\documentclass[aps,twocolumn,superscriptaddress,english,10pt,nofootinbib]{revtex4}

\usepackage{xcolor}
\usepackage{graphicx}
\usepackage{latexsym}
\usepackage{dcolumn}
\usepackage{bm}
\usepackage{slashed}

\usepackage{hyperref}
\usepackage{amsmath,amssymb,physics}
\usepackage{parskip}
\usepackage{adjustbox}
\usepackage{mathtools}
\usepackage{empheq}
\usepackage[normalem]{ulem}
\usepackage{verbatim}
\hypersetup{colorlinks=true, linkcolor=magenta, filecolor=cyan, urlcolor=blue}

\newlength\dlf

\newcommand{\mr}[0]{\mathrm}
\newcommand{\bo}[0]{\mathbf}
\newcommand{\rec}{\mathrm{{rec}}}
\newcommand{\tg}[0]{\Tilde{g}_{\phi\gamma}}

\begin{document}

\title{Stochastic Gravitational Waves from Narrow-Band Cosmological Magnetic Fields}

\author{Abdias Aires}
\email{abdias.aires@unesp.br}
\affiliation{Department of Physics, McGill University, Montreal, QC H3A 2T8, Canada}
\affiliation{Instituto de Fisica Teorica, UNESP-Universidade Estadual Paulista R.~Dr.~Bento~T.~Ferraz~271,~Bl.~II, Sao Paulo 01140-070, SP, Brazil}

\author{Tatsuya Daniel}
\email{tdaniel@physics.mcgill.ca}
\affiliation{Department of Physics, McGill University, Montreal, QC H3A 2T8, Canada}
\affiliation{Trottier Space Institute, McGill University, Montreal, QC H3A 2T8, Canada}

\begin{abstract}
We calculate the spectrum of gravitational waves generated by a cosmological magnetic field with a narrow-band power spectrum. For a magnetic spectrum narrowly peaked at a comoving wavenumber $k_c$, we find that the gravitational wave signal has a maximum at $2k_c$ and, as expected from momentum conservation, has a sharp cutoff at that scale. We apply this analysis to an ultralight dark matter model of magnetogenesis operating around recombination, in which an axion-photon coupling generates a helical magnetic field by parametric resonance. We determine the conditions for the resulting gravitational waves to enter the frequency range of pulsar timing experiments. Although the required frequency can be reached for sufficiently large coupling, the corresponding amplitude is below the experimental sensitivity for physically allowed energy densities of the magnetic field. We generalize this conclusion to models of magnetogenesis operating at other epochs, estimating as $T \simeq 160$ MeV the temperature of a magnetogenesis process assumed to be the source of the stochastic gravitational wave background suggested by pulsar timing arrays. We place the lower bound $T \gtrsim 1$-10 MeV on such viable models.
\end{abstract}

\date{November 17, 2026}

\maketitle

\section{Introduction}\label{sec:intro}
The existence of magnetic fields on cosmological scales has been suggested by indirect evidence from the late universe. The nonobservation of $\gamma$-ray cascade emissions indicates the lower bound $B(r) > 10^{-17}$ Gauss for scales $r$ larger than $10^{-1}$ Mpc \cite{Neronov2009,Grasso2001,Neronov2010,Durrer2013,Kahniashvili2013,Subramanian2016,Vachaspati2021}, while the magnetic fields observed in galaxies and galaxy clusters may have been amplified from primordial seeds~\cite{Kulsrud:1996km,Kulsrud:2007an,Brandenburg:2004jv,Widrow:2002ud}
(see~\cite{Grasso:2000wj,Kandus:2010nw,Durrer:2013pga,Subramanian:2015lua,Vachaspati:2020blt}
for reviews). These considerations motivate a variety of cosmological magnetogenesis models, which typically operate before Big Bang nucleosynthesis (BBN), e.g., during inflation or in phase transitions in the early universe \cite{Turner:1987bw,Sigl1997,Giovannini2008,Kahniashvili2010}. 

An important consequence of cosmological magnetic fields is the production of gravitational waves (GWs). The electromagnetic stress-energy tensor contains an anisotropic stress component which sources the transverse-traceless part of the metric perturbation. Consequently, a stochastic magnetic field generates a stochastic GW background (SGWB) \cite{Caprini2001,Caprini2004,Caprini2006}. Because the anisotropic stress is quadratic in the magnetic field, its power spectrum is determined by a convolution of the magnetic field power spectrum \cite{Durrer2000,Lewis2004,Paoletti2011}. The resulting GW spectrum therefore depends not only on the field strength but also on the spectral structure, and in general does not simply reproduce the magnetic spectrum itself.

This observation motivates a general procedure for connecting magnetogenesis models to stochastic GW signals. Given a magnetic field power spectrum $P_B(k)$ generated at some epoch, one can construct the power spectrum of its transverse-traceless anisotropic stress and then solve the sourced tensor equation to obtain the resulting GW spectrum. In this work, we develop this mapping explicitly for narrow-band magnetic spectra. In particular, for a magnetic spectrum narrowly concentrated around a comoving wavenumber $k_c$, we obtain an analytic expression for the convolution entering the GW spectrum. The resulting GW power has a maximum at $k = 2 k_c$ and, as expected from momentum conservation, has a sharp cutoff at that scale. The characteristic GW frequency is therefore related to, but is not identical to, the characteristic magnetic field frequency.



As a concrete application of this framework, we consider magnetogenesis from an axion-photon coupling. Axion-like fields coupled through an interaction of the form $\phi F\Tilde{F}$ can amplify the helicity modes of the electromagnetic field through a tachyonic instability during coherent axion evolution \cite{Barnaby2011,Barnaby2012}. Recently, this mechanism has been applied to ultralight dark matter (ULDM), with the axion beginning to oscillate around the epoch of recombination and generating helical magnetic fields on cosmological scales \cite{brandenberger2025}. Subsequent studies have investigated narrow-band resonance and backreaction effects in this scenario \cite{brahma2026}, as well as the production of secondary  Lyman-Werner photons assisting the formation of supermassive black holes by direct collapse in the early universe \cite{brandenberger2026, Aires_2026}. The resulting magnetic spectrum is concentrated around a characteristic comoving scale, making this mechanism a particularly useful application of the magnetic field-to-GW mapping developed here.

The nanohertz (nHz) frequency range of GWs is of particular interest, as several pulsar timing array (PTA) collaborations have reported evidence for a SGWB in that range \cite{Agazie2023,PPTA2023,EPTA2023,IPTA2022}. Although the observed signal is consistent with an astrophysical origin, cosmological sources remain of interest as possible explanations. For magnetogenesis models, this motivates the question of whether the GWs produced by their magnetic fields can be sufficiently strong and sufficiently low in frequency to be observable by PTAs.

We address this question in two complementary ways. First, we derive the GW spectrum sourced by a narrow-band magnetic field power spectrum. We apply these results to the ULDM magnetogenesis mechanism of \cite{brandenberger2025}, determining the coupling constant required to place the resulting GW signal in the PTA frequency range and relating the GW amplitude to the fraction of axion energy transferred into the gauge sector. We find that significant GW production can happen before the field decays by magnetohydrodynamic (MHD) turbulence. Although the characteristic frequency can be in the nHz range for sufficiently large coupling, the resulting GW amplitude remains below the sensitivity of current and projected PTA experiments for physically allowed energy densities of the magnetic field. 

Second, we use the analytical result for the GW spectrum to investigate how the efficiency of the GW production depends on the epoch of magnetogenesis. Expressing the efficiency in terms of the temperature at which the magnetic field is generated, we obtain a direct scaling of the GW amplitude with the temperature of magnetogenesis. We use this scaling to estimate the epoch at which magnetogenesis can efficiently source a PTA-band signal. In particular, for a representative frequency of $\sim10^{-8}\,\mathrm{Hz}$, we find that magnetogenesis models below temperatures of $\sim 1$-$10\,\mathrm{MeV}$ cannot generate sufficient GW power to explain PTA measurements under the assumptions of our calculation. We estimate as $T \simeq 160$ MeV the temperature of the magnetogenesis process assumed to be a source of the SGWB, placing it near the QCD phase transition scale.\footnote{The possibility of producing PTA-band GWs has also been investigated in previous work, e.g. \cite{RoperPol2022}.}

Throughout this paper, we set $c = \hbar = 1$ and use $M_{\text{pl}} = 1/\sqrt{8\pi G}$, unless otherwise noted, and we assume a $(-,+,+,+)$ metric signature; Greek letters ($\mu$,$\nu$,...) range over all spacetime coordinates, Latin letters (i,j,...) range over spatial indices, and derivatives take the form $\partial_\alpha = \frac{\partial}{\partial x^\alpha}$.

\section{ULDM and magnetogenesis at recombination}\label{sec:magnetogenesis}
\noindent 
We begin by reviewing the ULDM magnetogenesis scenario. The mechanism is based on the 
Lagrangian \cite{brandenberger2025}:

\begin{equation}
    \mathcal{L} = -\frac{1}{2}\partial^\mu\phi \partial_\mu\phi
    - V(\phi) - \frac{1}{4}F_{\mu\nu}F^{\mu\nu} 
    + g_{\phi\gamma}\phi F_{\mu\nu} \Tilde{F}^{\mu\nu},
\end{equation}
where $\phi$ is a pseudoscalar ULDM field with a potential $V(\phi)$, $F_{\mu\nu}$ is the field strength tensor of the vector potential $A_{\mu}$, $\Tilde{F}^{\mu\nu} \equiv \epsilon^{\mu\nu\rho\sigma}F_{\rho\sigma}/2$ is its dual, with $\epsilon^{\mu\nu\rho\sigma}$ as the 4D Levi-Civita tensor, and $g_{\phi\gamma}$ is a coupling constant.

We suppose the field $\phi$ to have been displaced from the minimum of its effective potential
by some misalignment mechanism
in the early universe. Its mass
is assumed to be small enough that the field starts oscillating ($H \sim m$) 
not long before recombination. Therefore, its oscillations are coherent
across cosmological scales by that epoch. 

The gauge field in this mechanism does not have to be the Standard Model electromagnetic field: any sufficiently light $U(1)$ gauge field coupled to the ultralight scalar can undergo the same instability and source GWs through its stress-energy tensor. We specialize here to the electromagnetic realization, for which the effects of the residual plasma conductivity require separate consideration.\footnote{In the electromagnetic realization of the ULDM model considered here, finite conductivity can possibly suppress the amplification of the gauge field and therefore modify the resulting magnetic field spectrum.
This effect depends on the time-scale of the resonance as compared to the mean time between collisions in the plasma.
The reader is referred to \cite{Sharma2026} and the appendix of \cite{brahma2026} for a discussion of residual ionization and conductivity effects.}

Assuming that $\phi$ depends
only on time, and setting the conductivity $\sigma$ to zero after
recombination, we obtain the following equation of motion for the
helicity modes of the gauge field:

\begin{equation}
\label{eq:hel_modes}
    (\partial^2 + k^2 \pm k g_{\phi\gamma}a(\eta)\dot\phi)\mathcal{A}_\pm = 0,
\end{equation}

\noindent where $k$ is the comoving wavenumber in an expanding background, $\eta$ is conformal time, and $\mathcal{A}_+$ and $\mathcal{A}_-$ are the two helical
modes. 

The effective frequency squared of each helicity mode can become negative for a sufficiently large axion-induced contribution. In particular, depending on the sign of $\dot{\phi}$, one of the two helicities becomes tachyonic when the magnitude of the axion-induced term exceeds the vacuum contribution $k^2$. During such intervals, the corresponding gauge-field mode undergoes exponential growth, resulting in efficient amplification over a narrow range of comoving wavenumbers. Because $a(\eta)$ is time-dependent and these intervals are different
for $\mathcal{A}_+$ and $\mathcal{A}_-$,
the effect on the helicity modes is asymmetric.
As a result, the magnetic field produced will be helical.

The rate of exponential growth is given by the Floquet exponent,

\begin{equation}
\label{eq:floquet}
    \mu_k = (g_{\phi\gamma} a m \phi(t)k)^{1/2}.
\end{equation}
The resonant effect therefore increases with $k$ until some critical wavenumber 
$k = k_c$, which is the maximum $k$-value for which \eqref{eq:hel_modes} can become tachyonic, depending
on the amplitude of $\Dot\phi$. The time variation of the pseudoscalar field can be related to the initial amplitude of oscillations at 
recombination as $\dot \phi \sim m\phi_0$, where $m$ is the mass of the field. Thus, the wavenumber at which the instability shuts off is:

\begin{equation}
\label{eq:kc}
    k_c(\eta) \simeq g_{\phi\gamma} m \phi_0 a(\eta).
\end{equation}

Assuming that most of the DM is composed of $\phi$ at
recombination (which gives $m\phi_0^2 \sim T^2_\rec$ \cite{brandenberger2025}),
the critical comoving wavenumber can be estimated 
as:

\begin{equation}
\label{eq:lambda_c}
    k_c \sim \Tilde{g}_{\phi\gamma} 10^{6} \, \mr{Mpc}^{-1},
\end{equation}

\noindent where $\Tilde{g}_{\phi\gamma}$ is defined as

\begin{equation}
    g_{\phi\gamma} \equiv \Tilde{g}_{\phi\gamma} 10^{-10} \, \mr{GeV}^{-1}.
\end{equation}
A conservative upper bound is 
\cite{Marsh_2016}:

\begin{equation}
    \Tilde{g}_{\phi\gamma} \ll 1.    
\end{equation}

\noindent Since the phase space scales as $k^3$, most of the energy flows
into modes with $k \sim k_c$. Therefore, the magnetic field $B(k)$ will have a peak at that wavenumber. 
Indeed, tachyonic resonance causes exponential amplification by a factor $e^{2\mu_kt}$, 
with $\mu_k$ given by \eqref{eq:floquet}. 

For a comparison across spatial scales, one can
note that the exponent at $k_c$ is larger than 
at $k_1 \equiv 10 \, \mr{Mpc}^{-1}$ by a factor:

\begin{equation}
    \frac{\mu_{k_c}}{\mu_{k_1}} \sim \tg^{1/2} 10^{2.5},
\end{equation}

\noindent using \eqref{eq:lambda_c}.
For $\tg \sim 0.1$, this gives a ratio of $\sim$100 between exponents. 
Hence, the field strength
at scales smaller than $10^{-1}$ Mpc is expected to be 
larger than at $10^{-1}$ Mpc by at least a few orders of 
magnitude.\footnote{The amplification will be limited
by backreactions. However, these are expected to be under control by the
estimates of \cite{brahma2026}.}

It follows that the power spectrum of the magnetic field for $k = k_c$ is
well approximated by the total energy density in the magnetic field:

\begin{equation}
    P_B(k_c) \sim \mathcal{F}T^4_\mr{rec},
\end{equation}

\noindent where $\mathcal{F}$ is the fraction of the 
energy density which has been drained from the axion field,
and $T_\mr{rec} \sim 0.1 $eV is the temperature of the universe at 
recombination.\footnote{The energy density of DM is $\sim T_\rec^4$ by
recombination.} The corresponding field strength is:

\begin{equation}
\label{eq:B_kc}
    B(k_c) \sim \mathcal{F}^{1/2} \text{ Gauss}.
\end{equation}
One expects the magnetic field to eventually decay by MHD turbulence. As we will show
in Sec.~\ref{sec:gw-spectrum}, GW production is expected 
to happen on a much shorter timescale. The resulting gravitational
spectrum will therefore have the imprint of the initial
magnetic field spectrum, highly-peaked at the critical wavenumber
$k_c$.

Additionally, the possibility of a narrow-band
parametric resonance in the equation of motion for the helicity
modes \eqref{eq:hel_modes} has been studied and has been shown to 
be more efficient than the process described above for certain
regions of parameter space \cite{brahma2026}. In this case, the resulting magnetic
field in this case has a sharp peak at the wavenumber:

\begin{equation}
    k_p \sim m_{20} 10^{6} \, \mr{Mpc}^{-1},
\end{equation}
with $m_{20}$ defined as:

\begin{equation}
    m \equiv m_{20} 10^{-20} \, \mr{eV},
\end{equation}

\noindent and $m_{20} \gtrsim 1$ \cite{brahma2026}. Since in both
cases the power spectrum is concentrated at a
particular wavenumber, we will refer to a ``narrow-band'' 
spectrum as the outcome of either process in what follows.








\section{GWs from a narrow-band magnetic field}\label{sec:gw-spectrum}
In this section, we calculate the power spectrum of GWs generated by a narrow-band magnetic field. 
In the presence of an anisotropic stress $\Pi_{ij}$ acting
as a source, the
gravitational tensor modes evolve according to:

\begin{equation}
\label{eq:eom_Caprini}
    h''_{ij} + 2 \frac{a'}{a} h'_{ij} + k^2 h_{ij} = 8\pi G \Pi_{ij},
\end{equation}

\noindent where $h_{ij}$ is the transverse and traceless metric 
perturbation on an FLRW background,
\begin{align}
    ds^2 &= a^2(\eta)\left[ -d\eta^2 + (\delta_{ij} + 2h_{ij}) dx^i dx^j \right], \\
    \partial^ih_{ij} &= 0, \quad h^i_{~i} = 0,
\end{align}

\noindent and primes denote derivatives with respect to conformal time $\eta$. 
We are interested in the case of a Gaussian, homogeneous, and isotropic
magnetic field.
As studied in detail in \cite{Caprini2001}, each component of $\Pi_{ij}$ is given by:

\begin{equation}\label{eq:Pi-component}
    \Pi_{\bullet}(k,\eta) = \frac{1}{a^2}f(k,\eta)
        \Tilde{\Pi}_\bullet(k),
\end{equation}

\noindent where $\Tilde\Pi_\bullet(k)$ is a 
time-independent random variable with power spectrum
$\langle |\Tilde\Pi_\bullet(k)|^2 \rangle = 1$. 
The function $f(k,\eta)$ will be related to the magnetic power 
spectrum $P_B(k)$. In the isotropic and homogeneous case, one
has:

\begin{equation}
\label{eq:PB_Caprini}
    \langle B^i(\bo k) B^{*j}(\bo k)\rangle \equiv
        (\delta^{ij} - \hat{k}_i \hat{k}_j ) 
        \delta^3(\bo k - \bo k') P_B(k).   
\end{equation}
Here, $B^i(\bo k)$ is a comoving field, related to the
physical field at time $\eta$ as $\Tilde{B}^i(\bo k,\eta) 
= B^i(\bo k)/a^3$.\footnote{The physical field therefore redshifts
as $\Tilde{B}^2 \sim a^{-4}$.} In the case of a time-independent
power spectrum, we have \cite{Caprini2001}:

\begin{equation}
\label{eq:f2B}
    f^2(k) = \frac{1}{16(2\pi)^8} 
    \int d^3q P_B(q) P_B(|\bo k - \bo q|) 
    (1 + 2\gamma^2 + \gamma^2\beta^2),
\end{equation}

\noindent with $\gamma = \hat{\bo k} \cdot \hat{\bo q}$
and $\beta = \hat{\bo k} \cdot \widehat{\bo k - \bo q}$. (The 
approximation of time-independence will be justified below.) 
From \eqref{eq:eom_Caprini}, it follows that each component of
the GW is given by:

\begin{equation}
    h_\bullet(k,\eta) = h(k,\eta) \Tilde \Pi_\bullet(k),
\end{equation}

\noindent where $h(k,\eta)$ is a solution to:

\begin{equation}
\label{eq:h_eom}
    h'' + 2\frac{a'}{a}h' + k^2 h 
        = \frac{8\pi G}{a^2(\eta)} f(k),
\end{equation}
using \eqref{eq:Pi-component}. We will apply the subhorizon limit 
to the rescaled variable $\Tilde{h} \equiv ah$, 
which satisfies:

\begin{equation}\label{eq:gw-rd-prior}
    \Tilde{h}'' + \left(k^2 - \frac{a''}{a}\right) \Tilde{h} = 
    \frac{8\pi G}{a}f(k).
\end{equation}

\noindent In radiation domination (RD), $a'' = 0$, while $a''/a = 2/\eta^2$ 
during matter domination (MD). For deeply subhorizon modes $k\eta \gg 1$,
the second term between
parentheses in \eqref{eq:gw-rd-prior} can be neglected. We can therefore write:

\begin{equation}
    \Tilde{h}'' + k^2\Tilde{h} = \frac{8\pi G}{a}f(k),
\end{equation}
which is valid for all modes in RD or for modes well-within the horizon in MD.
This equation corresponds to the Green's function:

\begin{equation}
\label{eq:Green}
    G_h(k;\eta,\tau) = \frac{1}{k} \sin k(\eta-\tau)
        \Theta(\eta-\tau),
\end{equation}
where $\Theta$ is the Heaviside step function. The solution is therefore:

\begin{equation}
\begin{aligned}
\label{eq:th}
    \Tilde{h}(k,\eta)
        &= 8\pi G 
        \int_{\eta_i}^\eta G_h(k;\eta,\tau) 
        \frac{f(k)}{a(\tau)} d\tau \\
        &= \frac{8\pi G}{k} f(k)
        \int_{\eta_i}^\eta 
        \frac{\sin k(\eta-\tau)}{a(\tau)} d\tau, \\ 
\end{aligned}    
\end{equation}

\noindent where $\eta_i$ is the generation time of the magnetic field. 
The time derivative of $\Tilde{h}(k,\eta)$ is:

\begin{equation}\label{eq:h'}
\begin{aligned}
    \Tilde{h}'(k,\eta) = 8\pi G f(k) \int^\eta_{\eta_i}
        \cos k(\eta-\tau) \frac{1}{a(\tau)}d\tau,
\end{aligned}
\end{equation}

\noindent and the GW spectrum is given as \cite{Caprini2001}:

\begin{equation}
\label{eq:omega_h}
    \frac{d \Omega_G(k)}{d \log k} \equiv 
        \frac{d \rho_G(k)}{\rho_c d \log (k)} 
        = \frac{k^3 {h'}^2}{a^2 \rho_c(2\pi)^6 G},
\end{equation}

\noindent where $\rho_c = 3 H_0^2/(8\pi G)$ is the
critical density today. From the definition $h =
\Tilde{h}/a$, we have:

\begin{equation}
\begin{aligned}
\label{eq:h'h}
    h' &= \frac{1}{a}\left( \Tilde{h}' - \Tilde{h} 
        \frac{a'}{a} \right) \\
        &= \frac{\Tilde{h}'}{a}(1 + \mathcal{O}[k_H(\eta)/k]),
\end{aligned}
\end{equation}

\noindent where $k_H \equiv 2\pi/R_H$ and $R_H \equiv a/a'$ is the comoving Hubble radius. The second equality follows from $\Tilde{h}' \sim k \Tilde{h}$, 
which is a consequence of \eqref{eq:th} and \eqref{eq:h'}. The condition $k_H(\eta)/k \ll 1$ holds at late times $\eta$ 
even if the mode $k$ is near the Hubble scale at generation $\eta_i$,
provided $\eta \gg \eta_i$. This is true for processes taking place
near recombination or during RD. 

From \eqref{eq:h'} and
\eqref{eq:h'h}, we can write $h'(k,\eta)$ as:

\begin{equation}
\label{eq:h'f}
\begin{aligned}
    h'(k,\eta) &= \frac{8\pi G f(k)}{a(\eta)} 
        \int^\eta_{\eta_i}
        \frac{\cos k(\eta-\tau)}{a(\tau)} d\tau\\
    &= \frac{8\pi G f(k)}{a a_i k} (\sin k(\eta-\eta_i) + \mathcal{O}[k_H(\eta_i)/k]), 
\end{aligned}
\end{equation}

\noindent where $k_H(\eta_i)/k$
is evaluated at the time of generation. This expression is valid for either RD or MD.  For the process described in \cite{brandenberger2025}, 
we have $k/k_H(\eta_\rec) \sim 3\tg \times 10^8
\gg 1$ if $\tg$ satisfies the lower bound for efficient resonance \cite{brandenberger2025}.\footnote{This bound is $\tg > 10^{-9} T/T_0$, where $T$ is the temperature during the resonance. For $T \simeq T_\rec$,
the condition is $\tg \gtrsim 10^{-6}$.} 
For magnetogenesis based on phase transitions, one typically has $k/k_H(\eta_i) \sim 
10^{1}$-$10^{4}$ \cite{Durrer:2008eom,Kahniashvili2013}. From \eqref{eq:omega_h}, this gives the spectrum:

\begin{equation}
\label{eq:omega_G1}
    \frac{d \Omega_G(k)}{d \log k} = 
        \frac{Gf^2(k)}{a^4 a_i^2 \rho_c \pi^4} k \sin^2k(\eta - \eta_i).
\end{equation}

\subsection{Narrow-band magnetic field spectrum}
\label{sec:peak_spec}

To obtain an expression for $f^2(k)$ from \eqref{eq:f2B}, 
we approximate $P_B(k)$ as highly-peaked at the
critical wavenumber $k_c$:

\begin{equation}
\label{eq:B_dirac}
    P_B(k) = \frac{2\pi^2 B_0^2}{k_c^2} \delta(k-k_c),
\end{equation}

\noindent where $B_0$ is a normalization constant to be derived shortly.
The sharp peak in the magnetic field spectrum follows from the generation
mechanism reviewed in Sec.~\ref{sec:magnetogenesis}. 

Causality implies a tail 
$P_B(k) = Ak^n$ for scales outside the Hubble radius at formation,
where $n \geq 2$ and $A$ is a constant. 
At recombination, one has again, $k/k_H(\eta_i) \sim 3\tg 
\times 10^8 \gg 1$. The large separation between $k_c$ and $k_H$, 
as well as the sharp peak predicted at $k_c$, allows one
to neglect the power on scales larger than the horizon when computing
$f^2(k)$.

We further approximate the spectrum as time-independent during the 
production of GWs, even though, as mentioned earlier, the magnetic
field is generally expected to decay by MHD turbulence in the plasma. The approximation of time-independence
can be justified as follows.
The timescale of GW production is $\delta t_\mr{GW} 
\sim 1/k_c$, as can be seen from the Green's function \eqref{eq:Green}. 
The timescale of turbulent decay is the eddy turnover time, $\delta t_e
= (v_A k_c)^{-1}$, where $v_A$ is the Alfvén speed \cite{Jedamzik1998,Subramanian2016}:

\begin{equation}
    v_A = \left[\frac{2 \rho_B}{\rho_b} \right]^{1/2} = 
    \left[ \frac{2 \mathcal{F} 
    \Omega_\mr{cdm}^{(0)}}{\Omega_b^{(0)}} \right]^{1/2}
    \ll 1,
\end{equation}

\noindent and $\rho_B$, $\rho_b$ are the magnetic field and baryon energy densities;  $\Omega_{\rm cdm}^{(0)}$ and $\Omega_b^{(0)}$ are the relative densities of cold dark matter (CDM) and baryons today. We have also used $\rho_B = \mathcal{F} \rho_\mr{cdm}$
for the energy density of the magnetic field at recombination. The ratio of
timescales is:

\begin{equation}
    \frac{\delta t_\mr{GW}}{\delta t_e } = v_A \ll 1,
\end{equation}
which justifies taking $P_B(k)$ (hence, $f^2(k)$)
as constant in time during GW production.

From \eqref{eq:f2B} and \eqref{eq:B_dirac}, we obtain:

\begin{equation}\label{eq:fsquared}
    f^2(k) = \frac{B_0^4}{(8\pi)^3} \frac{1}{k_c^2k}
    \left[ 1 + \frac{k^2}{2 k_c^2} + \left(\frac{k}{2k_c} 
    \right)^4 \right] \Theta(2k_c-k),
\end{equation}

\noindent so the gravitational power spectrum is therefore:

\begin{equation}
\begin{aligned}
\label{eq:grav_spec}
    \frac{d \Omega_G(k)}{d \log k} &= 
        \frac{G B_0^4}{512 \pi^7 a^4 a_i^2 \rho_c k_c^2}
        \left[ 1 + \frac{k^2}{2 k_c^2} + \left(\frac{k}{2k_c}
        \right)^4 \right] \\
        &\times\sin^2k(\eta-\eta_i) \Theta(2k_c-k).    
\end{aligned}
\end{equation}

\noindent The spectral shape differs from the power-law and broken-power-law gravitational
spectra obtained in models which assume MHD decay to have taken place before significant
GW production \cite{RoperPol2022}.


\subsection{Normalization of the magnetic field spectrum}

With the gravitational power spectrum laid out for a narrow-band magnetic field, here we connect the normalization constant
$B_0$ in \eqref{eq:B_dirac} to the fraction of energy $\mathcal{F}$
drawn from the axion field at recombination. In Gaussian units, the energy density in the magnetic field is

    \begin{equation}
    \begin{aligned}
    \label{eq:rho_B}
        \rho_B &= \frac{1}{8\pi} g_{ij} \langle \tilde{B}^i(\bo x,\eta)
            \tilde{B}^j(\bo x,\eta) \rangle \\
            &= \frac{1}{8\pi} a^{-4} \delta_{ij}
            \langle B^i(\bo x) B^j(\bo x) \rangle,
    \end{aligned}
    \end{equation}

\noindent where again, $\Tilde{B}^i(\bo x, \eta)$ is the physical field measured by an observer 
at time $\eta$. In Fourier space, this is:

    \begin{equation}
    \begin{aligned}
        \rho_B &= \frac{1}{(8\pi)(2\pi)^6} a^{-4} \delta_{ij}\nonumber \\ 
        &\times \int d^3k d^3k' \, \exp (-i \bo x \cdot \bo k)
        \exp (i \bo x \cdot \bo k') \langle B^i(\bo k) B^{j*} (\bo k') \rangle.
    \end{aligned}
    \end{equation}

    \noindent For a Gaussian isotropic field, the correlator in Fourier space is:

    \begin{equation}
    \begin{aligned}
         \langle B^i(\bo k) B^{j*} (\bo k') \rangle = \delta^3(\bo k - \bo k')
         (\delta^{ij} - \hat{k}^i \hat{k}^j) P_B(k),
    \end{aligned}
    \end{equation}

    \noindent and hence,

    \begin{equation}
        \rho_B = \frac{1}{256\pi^7 a^4} \int d^3k \, P_B(k).
    \end{equation}

    \noindent Using \eqref{eq:B_dirac}, this gives an energy density:

    \begin{equation}
        \rho_B = \frac{B_0^2}{32\pi^4a^4}.
    \end{equation}
    We set $\rho_B(a_\rec) = \mathcal{F}\rho_\mr{cdm}(a_\rec)$, 
    where $\rho_\mr{cdm}(a_\rec)$
    is the energy density of DM at recombination. This gives:

    \begin{equation}
    \label{eq:Bo_rhocdm}
        B_0^2 a_\rec^{-4} = 32\pi^4 \mathcal{F} \rho_\mr{cdm}(a_\rec),
    \end{equation}
    or $B_0 a_\rec^{-2} \simeq 91.7 \mathcal{F}^{1/2} \, \mr{Gauss}$.\footnote{Our conventions in the definition of the power spectrum
(amounting to prefactors $(2\pi)^3$) differ from those of \cite{brandenberger2025}, which leads to a difference of two orders
of magnitude in $B_0$. The results will be expressed in terms of
$\mathcal{F}$, which has the same meaning in the two contexts.}

    From \eqref{eq:rho_B}, one has $B_\mr{phys}(a) \equiv \sqrt{8\pi \rho_B}
    = B_0/(2\pi^{3/2}a^2)$. The parameter $B_0$ is therefore related to the 
    comoving magnetic field as:

    \begin{equation}
    \label{eq:B0_Bphys}
        B_0 = (2\pi^{3/2}) B^0_\mr{phys} \simeq 11.1 B^0_\mr{phys}.
    \end{equation}

\subsection{Gravitational signal of ULDM magnetogenesis}

From the results of subsection
\ref{sec:peak_spec}, we now calculate the gravitational signal of
the model introduced in \cite{brandenberger2025} and
reviewed in Sec.~\ref{sec:magnetogenesis}. 

Current and forthcoming PTA experiments will probe the frequency range
$\sim 1$-$100$ nHz \cite{Shannon2025}, and the gravitational spectrum \eqref{eq:grav_spec} has 
power at all wavenumbers $k \leq 2k_c$, with $k_c$ given in \eqref{eq:lambda_c}. The
frequency $\sim$1 nHz at observation corresponds to the comoving wavenumber 
$k = 6.5 \times 10^5 \, \mr{Mpc}^{-1}$. From \eqref{eq:lambda_c}, the condition
for minimum detectability in the PTA frequency range is therefore:
\begin{equation}
    \tg \gtrsim 0.3,
\end{equation}
\noindent which is consistent with the lower bounds on the axion-EM decay constant \cite{Ayala2014,CAST2017,ParticleDataGroup2024}.

The NANOGrav collaboration, as well as the International PTA (IPTA), Parkes PTA (PPTA) and European PTA (EPTA) 
collaborations, have reported evidence for a signal which, if interpreted as a SGWB, corresponds to an
amplitude $d \Omega_G/d \log k \sim 10^{-10}$-$10^{-9}$ at frequencies
of $\sim1$-5 nHz \cite{Agazie2023,PPTA2023,EPTA2023,IPTA2022}. For an
extra gravitational signal to be detected in the same frequency range, it must be
above that amplitude floor. The model considered in this work produces a gravitational spectrum with a 
maximum at $k = 2k_c$:

\begin{equation}
    \frac{d \Omega_G(2k_c)}{d \log k} = 
        \frac{G B_0^4}{256 \pi^7a^4 a_\mr{rec}^2 \rho_ck_c^2},
\end{equation}

\noindent where we have replaced the average $\langle \sin^2 k(\eta-\eta_i)\rangle
=1/2$ in \eqref{eq:grav_spec}. For $B_0$ normalized as in \eqref{eq:Bo_rhocdm}, we find:

\begin{equation}
\label{eq:amp_cond}
     \frac{d \Omega_G(2k_c)}{d \log k} 
     \lesssim 3 \times 10^{-21} \mathcal{F}^2.
\end{equation}

In the case of narrow-band resonance as studied in \cite{brahma2026}, the 
frequency requirement is, analogously,

\begin{equation}
    m_{20} \gtrsim 0.3.
\end{equation}

\noindent This is, however, superseded by the bound $m_{20} \gtrsim 1$ from
considerations of structure formation \cite{brandenberger2025,Irsic_2017}.
In this case, the strength of the signal decreases to:

\begin{equation}
     \frac{d \Omega_G(2k_c)}{d \log k}\bigg|_\text{narrow-band res.} 
     \lesssim 10^{-21} \mathcal{F}^2.
\end{equation}

\noindent For the allowed range
$\mathcal{F} < 1$, both signatures lie below the detectability range of PTA 
experiments\footnote{The SGWB would have
power in frequencies up to at least $\sim 10^{-8}$ Hz. If the same model 
is to be a source of the SGWB itself (rather than the source of an extra
signal in that range), the frequency condition will impose
$\tg \gtrsim 3$. This range
of values for $\tg$ is excluded by observational bounds \cite{Marsh_2016}.
In addition, the amplitude condition \eqref{eq:amp_cond} would be
stricter by two orders of magnitude.}, given the existence of signal at the level $\sim10^{-10}$-$10^{-9}$
in the same frequency range. Indeed, the
two signals differ in strength by a factor of $10^{-11} \mathcal{F}^2$, which is 
significantly below the uncertainty of current and forthcoming PTA measurements 
\cite{Agazie2023,PPTA2023,EPTA2023,IPTA2022}.

\section{A bound on the temperature of magnetogenesis 
as a source of the SGWB}\label{sec:magnetogenesis-temp}

As a further look into the GW spectra from magnetogenesis models, we can ask how early such mechanisms would generally need to take place in order to produce an observable PTA signal. To do this, we can invert Eq.~\eqref{eq:grav_spec} to provide an
estimate of the epoch of magnetogenesis as a source of the SGWB.
At the reference wavenumber $k_c$, one has: 

\begin{align}
\label{eq:Omega_G_num}
    &\frac{d \Omega_G(k_c)}{d \log k}\bigg|_\mr{today} = 
        3.8\times 10^{-10} \left(\frac{T_i}{100 \mr{MeV}}\right)^2 \nonumber \\ &\times
        \left(\frac{B_0}{4.8 \times 10^{-5} \, \mr{Gauss}}\right)^4
        \left( \frac{k_c}{1.94\times10^7 \,\mr{Mpc}^{-1}} \right)^{-2}.
\end{align}

\noindent The anchor $k_c = 1.94\times10^7 \,\mr{Mpc}^{-1}$
corresponds to a frequency of $3\times10^{-8}$ Hz at observation,
whereas $B_0 = 4.8 \times 10^{-5}$ Gauss
saturates the BBN bound, $\Omega_B = \rho_B/\rho_r \sim 0.1$ \cite{Grasso2001}
(\textit{cf}. \eqref{eq:B0_Bphys}). This corresponds to a comoving field 
of $5.3 \, \mu$G. 

The signal derived from PTA collaborations has the amplitude 
$d \Omega_G/d \log k \sim 10^{-9}$ at the frequency
$3\times 10^{-8}$ Hz \cite{Agazie2023,PPTA2023,EPTA2023,IPTA2022}. 
From \eqref{eq:Omega_G_num}, and for $B_0$ which saturates the BBN bound, we find the corresponding
temperature of magnetogenesis to be at the QCD scale,\footnote{The causal 
(Batchelor) tail, which is
valid for $k \lesssim k_H$, is close enough to the reference wavenumber $k_c$ to 
make the Dirac-delta approximation less reliable at $T = 160$ MeV, when
$k_c / k_H \simeq 1.5$-$1.7$. 
This stands in
contrast to the case at recombination, for which $k_c/k_H \sim 3\tg \times 10^8
\gg 1$.
However, since 
the convolution in \eqref{eq:f2B} produces a mixing of scales
in $P_B(k)$, one expects the order of magnitude of $d\Omega_G/d \log k$ 
not to be very sensitive to the shape of the spectrum, given the same total
power. The result \eqref{eq:T_ref} confirms this expectation.}

\begin{equation}
\label{eq:T_ref}
    T_i \simeq 160 \; \mr{MeV}.    
\end{equation}

\noindent This results confirms the
outcome of other analyses of magnetogenesis in the early universe 
\cite{RoperPol2022}.\footnote{Since 
$k_c / k_H \simeq 1.5$-$1.7$ at this temperature, the $\mathcal{O}(k_H/k_c)$ terms
in \eqref{eq:h'f} could be important. We have checked the
accuracy of \eqref{eq:omega_G1} to be within $3\%$ of a numerical
output for these scales.}

In addition, Eq.~\eqref{eq:grav_spec} disfavors the possibility of 
any causal model of magnetogenesis at temperatures $T \lesssim 10 \,\mr{MeV}$ to
be a 
source of the SGWB. Indeed, a bound on the temperature can be estimated from the amplitude of
the GW spectrum at its peak, $k=2k_c$:

\begin{equation}
\begin{aligned}
    \frac{d \Omega_G}{d \log k}\bigg|_\mr{max} &= 
        \frac{G B_0^4}{256 \pi^7a^4 a_i^2 \rho_ck_c^2} \\
        &=
        3.8\times 10^{-11} \, \Omega_{B,i}^2
        \left(\frac{g_*}{10.75}\right)^{-2/3} \\&
        \times\left(\frac{T_i}{1 \, \mathrm{MeV}}\right)^2
        \left(\frac{f}{3\times 10^{-8} \,\mr{Hz}}\right)^{-2},
\end{aligned}
\end{equation}

\noindent where $\Omega_{B,i} = \rho_B(t_i)/\rho_r (t_i)$, $t_i$ is 
the time of generation, and $g_*$ represents the effective number of relativistic degrees of freedom. This result is obtained in the 
extreme case in which the total power of the magnetic field is concentrated at the wavenumber $k_c$,
the field is assumed to be coherent on a timescale much larger than that
of GW production, and the GW spectrum, obtained from the convolution \eqref{eq:f2B},
is calculated at $2k_c$. We therefore consider
it a conservatively high estimate of the GW spectrum at that wavenumber, arguably
representing the maximum available GW power. We restrict
the analysis to $T \lesssim 10$ MeV so that the scale corresponding to $3\times 10^{-8}$ Hz
is well within the horizon at generation ($2k_c/k_H \gtrsim 10$). 

For a comparison with PTA results, we 
take $(d\Omega_G/d\log k)_\mr{PTA} = 10^{-10}/h^2$ as the reference amplitude 
at the frequency $3\times 10^{-8} \mr{Hz}$, where $h$ is the dimensionless Hubble parameter. This corresponds to the lower end of the
$2\sigma$ range obtained from a fit of power-law or broken-power-law
spectra to PTA data \cite{Afzal2023,Agazie2023}. We obtain:

\begin{equation}
\begin{aligned}
    &\frac{(d\Omega_G/d\log k)_\mr{max}}{(d\Omega_G/d\log k)_\mr{PTA}}
    \bigg|_{f=3\times10^{-8}\mr{Hz}} \\ &\quad \quad \lesssim
    \begin{cases}
      10^{-1}, & 1 \,\mr{MeV} \lesssim T \lesssim 10\, \mr{MeV}, \\
      10^{-3}, & T \lesssim 1 \, \mr{MeV}.
    \end{cases}     
\end{aligned}
\end{equation}

\noindent In the
range above the BBN scale ($  0.1\, \mr{MeV} \lesssim T_i \lesssim 10\, \mr{MeV}$), 
we have assumed the BBN upper bound on the
energy density of the magnetic field, $\Omega_{B,i} \lesssim 0.1$. This result indicates that
magnetogenesis models operating below $T_i \simeq 10\, \mr{MeV}$ cannot produce
sufficient GW power at the frequency $f=3\times10^{-8}\mr{Hz}$
to explain current PTA observations.

As a caveat to the above analysis, we note that the level 
$(d\Omega_G/d\log k)_\mr{PTA} = 10^{-10}/h^2$ at $f=3\times10^{-8}\,\mr{Hz}$ 
(at the lower end of the $2\sigma$ range)
has been inferred from power-law and 
broken-power-law fits to PTA data, not from a pointwise fit at some reference
frequency. The bound can be updated as other analyses become available.

\section{Conclusion}\label{sec:conclusion}
In this work, we have calculated the SGWB sourced by narrow-band cosmological magnetic fields. As a concrete application, we have considered the ULDM magnetogenesis mechanism from \cite{brandenberger2025} and \cite{brahma2026}. We have found that significant GW production can happen before the magnetic field decays by MHD turbulence. While the post-recombination ULDM mechanism can generate cosmologically interesting magnetic fields, the GWs sourced by their anisotropic stress do not provide an explanation for the stochastic background suggested by PTA measurements. 

We find a suppressed gravitational signal to be a general property of magnetogenesis models not operating in the very early universe. In particular, we estimate as $T \simeq 160$ MeV 
the temperature of the magnetogenesis process assumed to be a source of the SGWB, given the amplitude of the background suggested by recent PTA measurements. Interpreting the predicted signal as a conservative high estimate of the GW power spectrum at a reference frequency, we place the lower bound $T \gtrsim 1$-10 MeV on the temperature of such viable models.

As a suggestion for future work, we note that the analysis of GW production can include the time-dependence of the magnetic power spectrum produced at recombination, i.e., one can investigate how its slow decay by MHD turbulence will affect the final gravitational spectrum. We note that this model offers an instance of cosmological magnetogenesis in which the generating mechanism is specified to good precision, so the assumption of a broken-power-law (post-MHD) spectrum doesn't need to be made as the initial condition for GW production.

We also note the possibility of improving the bound $T \gtrsim 1$-10 MeV. This can be done by 
comparing the predicted GW spectrum to the amplitude range obtained from a pointwise fit to PTA data at a viable reference frequency. 

\acknowledgments
The authors would like to thank Robert Brandenberger for useful discussions, as well as comments on a draft of this work. AA thanks the Department of Physics at McGill University for its hospitality. His work is supported by funds from CAPES - 001 and CNPQ, grant number 165240/2023-7. TD is supported by a Trottier Space Institute fellowship and by funds from NSERC.
\clearpage
\bibliography{bib}

\end{document}